\documentclass[structabstract]{aa}  
\usepackage{graphicx}
\usepackage{txfonts}
\newcommand{\HIIR}{H{\sc ii} region}
\begin{document}
   \title{Smoothed Particle Hydrodynamics simulations\\ of expanding {\HIIR}s}

   \subtitle{I. Numerical method and applications}

   \author{T. G. Bisbas
          \inst{1}
          ,
          R. W\"unsch
          \inst{1}
          , A. P. Whitworth,
	  \inst{1}
	  \and
	  D. A. Hubber
	  \inst{1,2,3}   
          }

   \institute{School of Physics and Astronomy, Cardiff University,
              Queens Buildings, The Parade, Cardiff, CF24 3AA, United Kingdom\\
              \email{thomas.bisbas@astro.cf.ac.uk}
         \and
             Institute for theoretical astrophysics, University of Oslo, Pb 1029 Blindern, 0315 Oslo, Norway
 	 \and
	    Centre of Mathematics for Applications, University  of Oslo, Pb 1053 Blindern, 0316 Oslo, Norway
             }

   \date{Received -; accepted -}

 
  \abstract
{Ionizing radiation plays a crucial role in star formation at all epochs. In contemporary star formation, ionization abruptly raises the pressure by more than three orders of magnitude; the temperature increases from $\sim 10\,{\rm K}$ to $\sim 10^4\,{\rm K}$, and the mean molecular weight decreases by a factor of more than 3. This may result in positive feedback (either by compressing pre-existing clouds and rendering them unstable or by sweeping up gravitationally unstable shells). It may also result in negative feedback (by dispersing residual dense gas). Ionizing radiation from OB stars is also routinely invoked as a means of injecting kinetic energy into the interstellar medium and as a driver of sequential self-propagating star formation in galaxies.}
{We describe a new algorithm for including the dynamical effects of ionizing radiation in SPH simulations, and we present several examples of how the algorithm can be applied to problems in star formation.}
{We use the HEALPix software to tessellate the sky and to solve the equation of ionization equilibrium along a ray towards each of the resulting tesserae. We exploit the hierarchical nature of HEALPix to make the algorithm adaptive, so that fine angular resolution is invoked only where it is needed, and the computational cost is kept low.}
{We present simulations of (i) the spherically symmetric expansion of an {\HIIR} inside a uniform-density, non--self-gravitating cloud; (ii) the spherically symmetric expansion of an {\HIIR} inside a uniform-density, self-gravitating cloud; (iii) the expansion of an off-centre {\HIIR} inside a uniform-density, non--self-gravitating cloud, resulting in rocket acceleration and dispersal of the cloud; and (iv) radiatively driven compression and ablation of a core overrun by an {\HIIR}.}
{The new algorithm provides the means to explore and evaluate the role of ionizing radiation in regulating the efficiency and statistics of star formation.}

   \keywords{Methods: numerical  --  Radiative transfer -- (ISM:) HII regions
               }

   \titlerunning{SPH simulations of expanding {\HIIR}s} \maketitle
%

\section{Introduction}

The influence of the UV radiation emitted by a massive star (or small group of massive stars) on the ambient interstellar medium is of great interest. As this radiation ionizes the surrounding neutral material, it creates a region of hot ionized gas (Str{\o}mgren 1939), known as an {\HIIR}. The huge contrast in pressure between the ionized gas in the {\HIIR} and the surrounding neutral gas causes the {\HIIR} to expand and sweep up the neutral gas into a dense shell (Kahn 1954; Spitzer 1978). This dense shell is likely to be prone to many different instabilities (e.g. Elmegreen 1994; Vishniac 1983), and some of these instabilities may trigger fragmentation and star formation. This mode of star formation is usually referred to as the ``collect and collapse'' mechanism (Elmegreen \& Lada 1977; Whitworth et al. 1994a,b). Deharveng et al. (2003, 2005, 2008) and Zavagno et al. (2007) have observed several instances where this mechanism appears to be operating.

The structure of the interstellar medium is observed to be extremely irregular and to contain many clouds. As an {\HIIR} expands, it may overrun and compress nearby clouds, causing them to collapse. This mechanism is known as ``radiation driven compression'', and has been simulated by several workers. Some of these simulations (Sandford et al. 1982; Bertoldi 1989; Lefloch \& Lazareff 1994; Miao et al. 2006; Henney et al. 2008) concentrate on the morphology of the resulting bright-rimmed clouds. In contrast, Peters et al. (2008) address the possibility that the flow of ionized gas off an irradiated cloud might be an important source of turbulence; and Kessel-Deynet \& Burkert (2003), Gritschneder et al. (2008), Dale et al. (2005) and Miao et al. (2008) explore the possibility that the collapse of a bright-rimmed cloud might sometimes lead to triggered star formation. These simulations show that it is possible to reproduce the observed morphologies of bright-rimmed clouds (Lefloch \& Lazareff 1995; Lefloch et al. 1997; Morgan et al. 2008). However, the evolution of bright-rimmed clouds and the role of expanding {\HIIR}s in triggering star formation are still poorly understood. For example, Dale et al. (2007) argue that the main effect of an expanding {\HIIR} may simply be to expose stars that would have formed anyway.

A variety of techniques have been used to follow the propagation of ionizing radiation in hydrodynamic simulations. Kessel-Deynet \& Burkert (2000) consider each SPH particle to be at the end of an individual ray, and compute the attenuation of ionizing radiation along this ray in order to determine whether the particle at the end is ionized. Dale et al. (2005, 2007) have refined this method so as to make it less compute-intensive. Alvarez et al. (2006) have invoked the HEALPix{\footnote{http://healpix.jpl.nasa.gov}} algorithm in order to distribute rays from the ionizing source in a regular manner. HEALPix (G\'orski et al. 2005) is a versatile tree structure which distributes pixels over the celestial sphere, to a user prescribed level of refinement. At each successive level of refinement a pixel from the level above is split into four smaller pixels. Each pixel corresponds to the end of a ray and each ray at a given level represents an equal solid angle on the celestial sphere. Abel \& Wandelt (2002) have shown how the HEALPix rays can be split adaptively, to increase the level of refinement where this is needed. This adaptive ray splitting has been used by Abel et al. (2007) and Krumholz et al. (2007).

In this paper we simulate the expansion of an {\HIIR} using the Smoothed Particle Hydrodynamics (SPH) code SEREN (Hubber et al., in preparation). SPH (Lucy 1977; Gingold \& Monaghan 1977) is a Lagrangian numerical method that describes a fluid using particles. The SEREN code is described in Section 3. The transport of ionizing radiation uses the hierarchical ray structure of HEALPix, with rays emanating from the ionizing source into the surrounding medium. We walk along each ray with an adaptive step proportional to the local smoothing length. At each step we calculate the attenuation of ionizing radiation by computing the number of recombinations into excited states of H$^{\rm o}$, i.e. we adopt the \textit{on-the-spot} approximation (Osterbrock 1974). Rays are split wherever the separation between particles becomes less than the separation between rays. This is a robust criterion, which ensures that a ray is split when it encounters a dense region, whilst a neighbouring ray passing through a more rarefied region is not split. Once the ionization front is located along a ray, the propagation of that ray is terminated, in order to reduce the computational overhead; this is particularly important during the early stages of evolution, when the Str{\o}mgren sphere only involves a small fraction of the total number of SPH particles. 

Although heavy elements play an important role in determining the temperature of interstellar gas, we do not consider their chemistry here. Instead, we assume that the composition of the gas is $X=0.7$ hydrogen and $Y=0.3$ helium, by mass; and that the helium is everywhere neutral. In the neutral gas we assume that the temperature is $T_{\rm n}=10\,{\rm K}$ and the hydrogen is all molecular; hence the mean molecular weight is $\mu_{\rm n}=2.35$ and the isothermal sound speed is $c_{\rm n}=0.2\,{\rm km}\,{\rm s}^{-1}$. In the ionized gas we assume that the temperature is $T_{\rm i}=10^4\,{\rm K}$ and the hydrogen is all ionized; hence $\mu_{\rm i}=0.678$ and $c_{\rm i}=11\,{\rm km}\,{\rm s}^{-1}$. In the transition region between the molecular and ionized regions, we impose a linear temperature gradient between these two limiting values.

We explore three cases. (i) In the first case, the star lies at the centre of a spherical cloud, and the {\HIIR} is spherically symmetric throughout its expansion. (ii) In the second case, the star is placed off-centre inside a spherical cloud; here the {\HIIR} breaks out of the cloud on one side, and the remainder of the cloud is accelerated by the rocket effect. (iii) In the third case, the star is located outside the cloud from the outset, and a shock is driven into the cloud ahead of the ionization front (i.e. radiation driven compression). These cases are presented only as illustrative examples of what the code can simulate. Detailed investigations of these phenomena will be presented in subsequent papers.

The paper is organized as follows. In Section \ref{sec.physics} we discuss briefly the expansion of an {\HIIR}, once the initial Str{\o}mgren sphere has formed. In Section \ref{sec.numerical} we describe in detail how we treat the propagation of ionizing radiation. In Section \ref{sec.applications} we test the algorithm on the three cases described above, and in Section \ref{sec.discussion} we discuss the results and conclude.


\section{The $D$-type expansion of an \HIIR}
\label{sec.physics}

\subsection{Spitzer solution}

Consider an arbitrary density field, $\rho({\bf r})$, and suppose that there is an ionizing star at the centre of co-ordinates. Assuming ionization equilibrium, and neglecting the diffuse radiation field, the position of the ionization front (IF), in the direction of the unit vector $\bf{\hat e}\,$, is given by ${\bf R}_{_{\rm IF}}=R_{_{\rm IF}}{\bf{\hat e}}\,$, where $R_{_{\rm IF}}$ is obtained from
\begin{equation}\label{integral}
\int_{r=0}^{r=R_{_{\rm IF}}}\rho^2\left(r{\bf{\hat e}}\right)r^2{\rm d}r=\frac{m^2\dot{\cal N}_{_{\rm LyC}}}{4\pi\alpha_{_{\rm B}}}\equiv I_{_{\rm MAX}}.
\end{equation}
Here, $m=m_{\rm p}/X=2.4\times 10^{-24}\,{\rm g}$ is the mass associated with one hydrogen nucleus when account is taken of the contribution from helium, $m_{\rm p}$ is the proton mass, $\dot{\cal N}_{_{\rm LyC}}$, is the rate at which the exciting star emits Lyman continuum photons, and $\alpha_{_{\rm B}}$ is the recombination coefficient into excited states only. Eqn. (\ref{integral}) assumes that the material inside the {\HIIR} is fully ionized.

Thus Eqn. (\ref{integral}) determines the radius $R_{_{\rm IF}}$ at which all the ionizing photons emitted in the direction $\bf{\hat e}$ have been used up balancing recombinations into excited states. We ignore recombinations straight into the ground state by invoking the {\it on-the-spot} approximation (Osterbrock 1974). 

Str{\o}mgren (1939) was the first to show that the transition from a state of almost completely ionized material to a state of almost completely neutral material occurs in a very short distance compared with the dimensions of the \HIIR. For example, for a spherically-symmetric {\HIIR} expanding into a cloud of uniform density $\rho_{\rm n}$, the radius of the ionization front is given by the Str{\o}mgren radius
\begin{eqnarray}\nonumber
R_{_{\rm St}}&\!=\!&\left(\frac{3m^2\dot{\cal N}_{_{\rm LyC}}}{4\pi\alpha_{_{\rm B}}\rho_{\rm n}^2}\right)^{1/3}\simeq 0.3\,{\rm pc}\,\left(\frac{\dot{\cal N}_{_{\rm LyC}}}{10^{49}\,{\rm s}^{-1}}\right)^{1/3}\left(\frac{\rho_{\rm n}}{10^{-20}\,{\rm g\,cm^{-3}}}\right)^{-2/3};\\\label{eq.stradius}
\end{eqnarray}
and the distance over which the degree of ionization changes from $90\,\%$ to $10\,\%$ is given by
\begin{eqnarray}\label{eqn.ifthick}
\Delta R_{_{\rm St}}&\simeq &\frac{20\,m}{\rho_{\rm n}\,\bar\sigma}\;\simeq\;0.0002\,{\rm pc}\left(\frac{\rho_{\rm n}}{10^{-20}\,{\rm g\,cm^{-3}}}\right)^{-1}\,.
\end{eqnarray}
Here $\bar\sigma=7\times10^{-18}\,{\rm cm}^2$ is the mean photoionization cross section presented by a hydrogen atom to Lyman continuum photons from an OB star.

Because the squared sound speed in the ionized gas inside the {\HIIR} is $c_{\rm i}^2\simeq122\,{\rm km}^2\,{\rm s}^{-2}$, whereas the squared sound speed in the neutral material outside the {\HIIR} is $c_{\rm n}^2\simeq 0.0352\,{\rm km}^2\,{\rm s}^{-2}$, there is a large pressure difference between the two regimes (more than three orders of magnitude), and this results in rapid expansion of the {\HIIR}. The outward propagation of the ionization front is subsonic relative to the ionized gas (where the sound speed is $c_{\rm i}\sim 11\,{\rm km\,s^{-1}}$), but supersonic relative to the neutral gas (where the sound speed is $c_{\rm n}\sim 0.2\,{\rm km}\,{\rm s}^{-1}$). Consequently a strong shock front appears ahead the ionization front (for a detailed discussion see Kahn 1954). Spitzer (1978) has obtained an approximate analytic solution for this phase of the evolution. In this solution, the radius of the ionization front is given by
\begin{eqnarray}\label{eq.spitzersolution}
R_{_{\rm IF}}(t)&=&R_{_{\rm St}}\left(1+\frac{7}{4}\frac{c_{\rm i}t}{R_{_{\rm St}}}\right)^{4/7}\,;
\end{eqnarray}
the density of the ionized gas by
\begin{eqnarray}\label{eq.solutionhiidensity}
\rho_{\rm i}=\rho_{\rm n}\,\left(1+\frac{7}{4}\frac{c_{\rm i}t}{R_{_{\rm St}}}\right)^{-6/7}\,;
\end{eqnarray}
the mass of ionized gas by
\begin{eqnarray}\label{eq.solutionhiimass}
M_{\rm i}(t)&=&\frac{m^2\,\dot{\cal N}_{_{\rm LyC}}}{\alpha_{_{\rm B}}\,\rho_{\rm n}}\,\left(1+\frac{7}{4}\frac{c_{\rm i}t}{R_{_{\rm St}}}\right)^{6/7}\,;
\end{eqnarray}
and the speed at which the ionization front propagates by
\begin{eqnarray}\label{eq.speedsolution}
\dot{R}_{_{\rm IF}}=c_{\rm i}\left(1+\frac{7}{4}\frac{c_{\rm i}t}{R_{_{St}}}\right)^{-3/7}\,.
\end{eqnarray}

\subsection{Semi-analytic approach}\label{semi-analytic}

A similar result can be obtained by solving the equation of motion of the accreting thin shell accelerated by the thermal pressure of the {\HIIR} (see e.q. Hosokawa \& Inutsuka 2006). We use this approach to obtain a more general solution which includes effects of the gravity and use this solution to calibrate our numerical code with the self-gravity switched on (see \S\ref{sec:grav}).

The equation of motion of the shell of mass $M_\mathrm{sh}$ and radius $R_\mathrm{sh}$ is
\begin{equation}
\label{semi_eom}
\frac{d}{dt}(M_\mathrm{sh}\dot{R}_\mathrm{sh}) = 4 \pi R_\mathrm{sh}^2 P_{\rm i}
 - \frac{G M_\mathrm{sh} M_{\rm i}}{R_\mathrm{sh}^2}
 - \frac{G M_\mathrm{sh}^2}{2R_\mathrm{sh}^2}
\end{equation}
where $P_{\rm i} = \rho_{\rm i} c_{\rm i}^2$, $M_{\rm i} = (4\pi R_\mathrm{sh}^3 \rho_{\rm i})/3$ and $\rho_{\rm i} = \left(\frac{3m^2\dot{\cal N}_\mathrm{LyC}}{4\pi\alpha_\mathrm{B}R_\mathrm{sh}^3}\right)^{1/2}$ are the pressure, the mass and the density of the ionized gas, respectively. The second term on the rhs represents the gravitational force due to the {\HIIR} mass, and the third term is the average gravitational force acting on the shell due to its own mass (it is zero at its inner edge and $GM_\mathrm{sh}^2 / R_\mathrm{sh}^2$ at the outer one; see Whitworth \& Francis 2002).

The mass of the shell is a difference between the mass of the original neutral gas which occupied the sphere of the radius $R_\mathrm{sh}$ and the mass of the {\HIIR}
\begin{equation}
\label{semi_Msh}
M_\mathrm{sh} = \frac{4\pi}{3} R_\mathrm{sh}^3 (\rho_{\rm n} - \rho_{\rm i}) \ .
\end{equation}
Inserting (\ref{semi_Msh}) into (\ref{semi_eom}) and using Eqn. (\ref{eq.stradius}) we get

\begin{eqnarray}
\label{semi_ode}
\ddot{R}_\mathrm{sh}(R_\mathrm{sh}^3 - R_\mathrm{St}^{3/2}R_\mathrm{sh}^{3/2})
+ 3\dot{R}_\mathrm{sh}^2 (R_\mathrm{sh}^2 -
R_\mathrm{St}^{3/2}R_\mathrm{sh}^{1/2}) & & \nonumber\\
= 3c_{\rm i}R_\mathrm{St}^{3/2}R_\mathrm{sh}^{1/2} 
- \frac{2\pi}{3} G\rho_{\rm n} R_\mathrm{sh}^4\left[ 1 -
  \left(\frac{R_\mathrm{St}}{R_\mathrm{sh}} \right)^3 \right] & & \ .
\end{eqnarray}
We solve this equation numerically using the subroutine {\tt integrate.odeint} from the Scientific python library (Jones et al. 2001). Since the model assumes an infinitesimally thin shell, the shell radius $R_\mathrm{sh}$ can be compared to the average radius of the ionization front and the average radius of the shock front $R_\mathrm{SF}$ in the numerical simulations (see Fig. \ref{fig.ifsf} and related discussion in \S\ref{sec:grav}).

\section{Numerical treatment}
\label{sec.numerical}

We use the SPH code SEREN which has been extensively tested using a wide range of standard and non-standard problems.  SEREN is an astrophysical SPH code parallelized using OpenMP and designed primarily for star and planet formation problems.  It contains both the traditional SPH method (Monaghan 1992) and the more recent `grad-h' SPH formulation (Price \& Monaghan 2004). There exists the option to include additional features within the basic SPH algorithm, such as time-dependent and/or Balsara-switched viscosity (Morris \& Monaghan 1997; Balsara 1995), artificial conductivity (Price 2008) and a variety of equations of state. The SPH equations can be solved with a choice of integrators, such as the symplectic 2nd order Leapfrog integrator, in conjunction with a block time-stepping scheme.  

SEREN uses a Barnes-Hut octal-spatial tree (Barnes \& Hut 1986), both to speed up the calculation of gravitational forces, and to facilitate the collation of particle neighbour lists.  The Barnes-Hut tree calculates the gravitational force up to quadrupole or octupole order, and has the option of using either the traditional geometrical opening angle criterion or the GADGET-style multipole acceptance criterion (Springel et al. 2001).  For problems where multiple gravitationally bound objects form, SEREN uses sink particles (Bate et al. 1995) to follow the simulation beyond the formation of the first dense bound objects.  A 4th-order Hermite integrator (Makino \& Aarseth 1992) is included, so that the code can follow the later N-body evolution of the sinks, after the accretion phase has finished. A paper describing SEREN in detail, and the tests to which it has been subjected, will be submitted shortly (Hubber et al., in preparation).

\subsection{Ray casting}

In order to determine the overall shape of the ionization front we consider many different directions $\bf{\hat e}\,$. These directions are chosen using the HEALPix algorithm (G\'orski et al. 2005). HEALPix generates a set of $N_{\ell}=12\times4^{\ell}$ directions (hereafter `rays'), where $\ell$ is the level of refinement and $\ell=0$ is the lowest level. The rays of level $\ell$ are distributed uniformly over the celestial sphere, as seen from the ionizing star. Each ray on level $\ell$ intersects the celestial sphere at the centre of an approximately square element of solid angle 
\begin{eqnarray}
\Delta\Omega_{\ell}=\frac{4\pi}{N_{\ell}}\,\,{\rm steradians}=\frac{\pi}{3\times4^{\ell}}\,\,{\rm steradians}. \nonumber
\end{eqnarray}
Therefore, the angle between neighbouring rays is of order 
\begin{eqnarray}
\Delta\theta_{\ell}\simeq\left(\Delta\Omega_{\ell}\right)^{1/2}\simeq 2^{-\ell}\,{\rm radians}.\nonumber
\end{eqnarray}

When HEALPix goes to a higher level, $\ell$, the angular resolution is increased by splitting each of the rays at the previous level, $\ell-1$, into four rays (see \S\ref{raysplitting}). In the simulations presented here, we limit the ray splitting to $\ell=7$, so there is a maximum of 196,608 rays at the ionization front. 

\begin{figure*}
\centering
\includegraphics[width=0.7\textwidth]{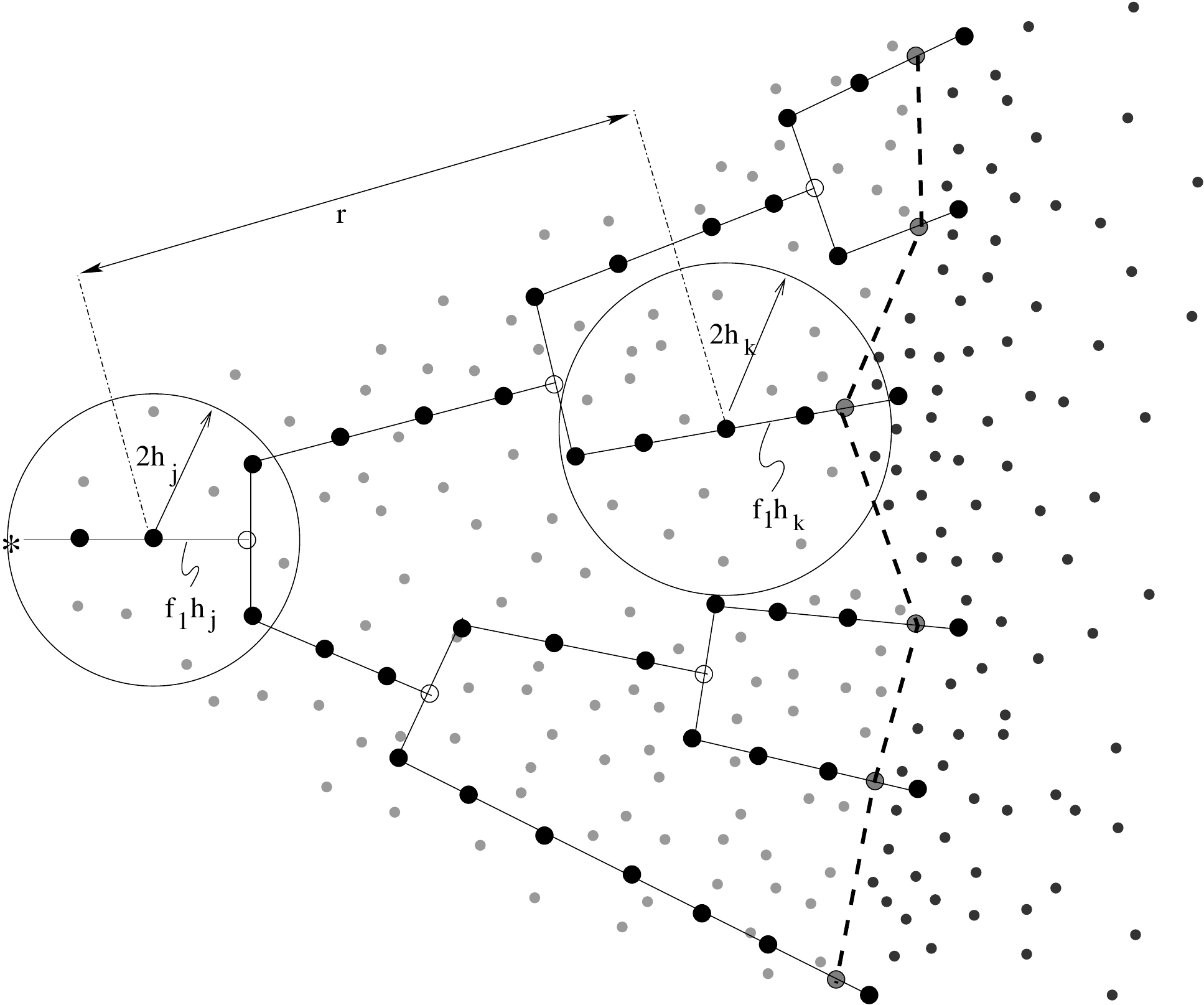}
\caption{ This figure shows the evaluation points (big black dots) along a family of rays (solid lines), starting from the ionizing star. To avoid confusion, the ray casting is plotted in 2-D. A ray is split as soon as its linear separation from neighbouring rays, $r_j\Delta\theta_{\ell}$, exceeds $f_{_2}h_j$, where $h_j$ is the smoothing length of the local evaluation point. Evaluation points where rays are split have been marked with open circles to indicate that they make no contribution to the summation in Eqn. (\ref{sumint}). A binary chop subroutine is used to locate the position of the ionization front (see \S\ref{locationif}).}
\label{fig: evalpoints}
\end{figure*}

The integral on the left-hand side of Eqn. (\ref{integral}) must be evaluated numerically along each ray until it equals $I_{_{\rm MAX}}$. To perform this integration we define a set of discrete evaluation points along each ray. At each evaluation point, $j$, we determine the density $\rho_j$ using an SPH summation over the ${\cal N}_{_{\rm NEIB}}$ nearest SPH particles, $k$,
\begin{eqnarray}
\rho_{j}=\sum_{k}\left\{\frac{m_{k}}{h^3_{j}}W\left(\frac{|{\bf r}_{k}-{\bf r}_{j}|}{h_{j}}\right)\right\}\,.
\end{eqnarray}
The integral in Eqn. (\ref{integral}) is then evaluated using the trapezium method
\begin{eqnarray}\nonumber
I(r_j)\,&=&\,\int_{r=0}^{r=r_j}\rho^2\left(r{\bf{\hat e}}\right)r^2{\rm d}r\\&\simeq &\sum_{j'=1}^{j'=j}\left\{\left(\frac{\rho^2_{j'-1}r^2_{j'-1}+\rho^2_{j'}r^2_{j'}}{2}\right)f_1h_{j'-1}\right\}\,,\\\label{sumint}
\end{eqnarray}
where $f_1$ is a dimensionless tolerance parameter of order unity controlling the integration step, and $h_{j'}$ is the smoothing length of evaluation point $j'$. The identifier $j'=0$ denotes the evaluation point placed at the position of the star, and consecutive evaluation points are then located according to
\begin{eqnarray}\label{epdis}
{\bf r}_{j'+1}={\bf r}_{j'}+f_1h_{j'}{\bf{\hat e}}\,.
\end{eqnarray}
Thus $f_1h_{j'}$ represents the adaptive step along the line of sight from the ionizing star towards the ionization front. Fig. \ref{fig: evalpoints} shows a representation of the method. Acceptable accuracy is obtained with $f_1=0.25$ and this is the value we use in the simulations presented here.

\subsection{Ray splitting}
\label{raysplitting}

The use of ray splitting allows us to maintain good resolution while achieving significant speed-up in comparison with uniform ray tracing (cf. Abel \& Wandelt 2002). In our scheme a ray is split into four child-rays as soon as its linear separation from neighbouring rays, $r_j\Delta\theta_{\ell}$, exceeds $f_2h_j$. Here $r_j$ is the distance from the ionizing star to evaluation point $j$, $f_2$ is a dimensionless parameter controlling the angular resolution of the ensemble of rays representing the ionizing radiation, and $h_j$ is the smoothing length of the local evaluation point $j$. Hence we require
\begin{equation}
\ell\geq\log_2\left(\frac{r_j}{f_2h_j}\right)\,.
\end{equation}
Acceptable results are obtained with $f_2$ in the range $(1.0,1.3)$. A smaller $f_2$ gives greater accuracy, but at the expense of more HEALPix rays. As one moves outwards away from the central star, rays are only ever split, they are never merged.

\begin{figure}[h]
   \centering
   \includegraphics[width=0.26\textwidth]{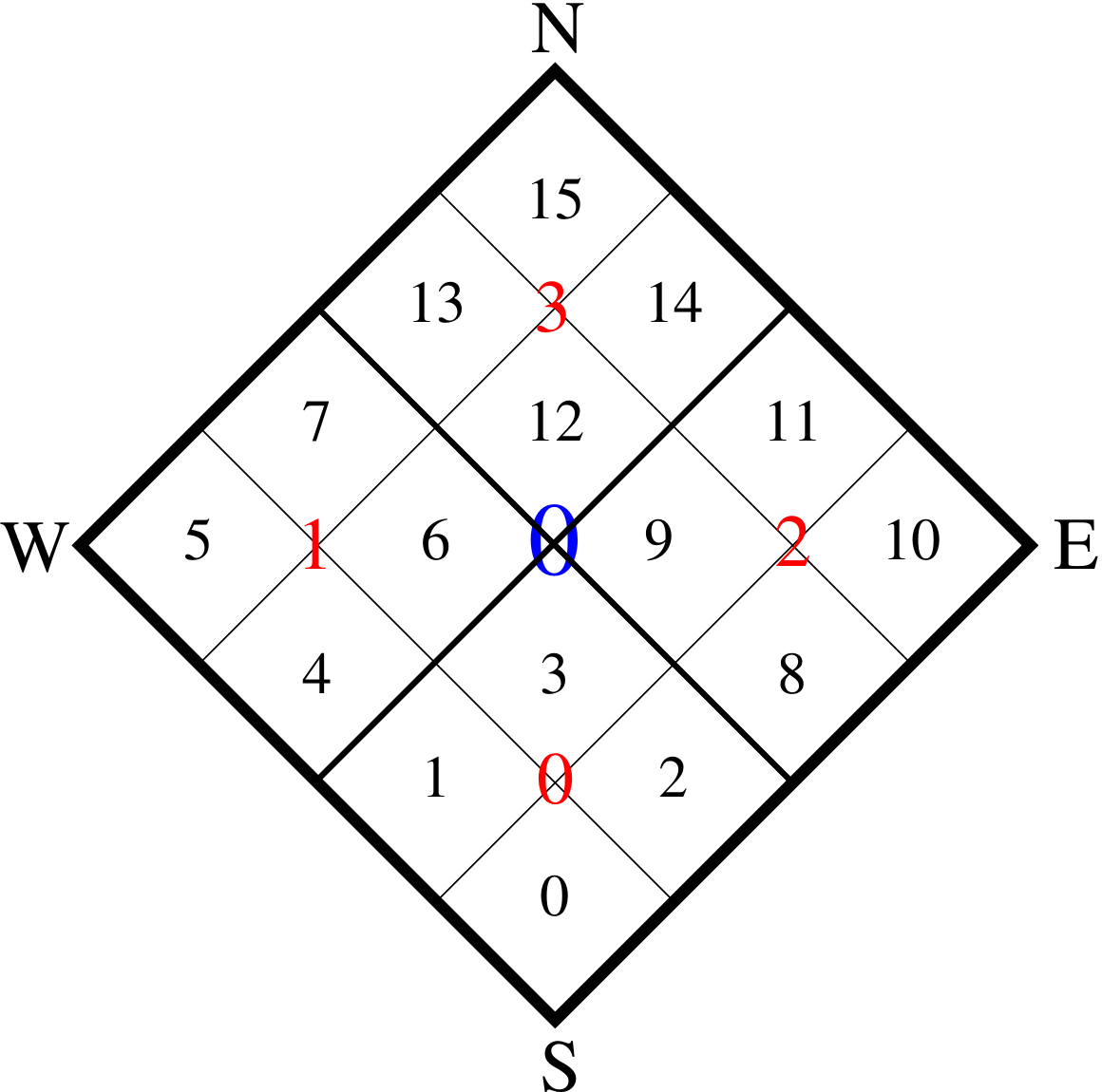}
\caption{Illustration of the scheme used for splitting a ray using HEALPix's \texttt{NESTED} scheme. This map shows the tessellation of the celestial sphere corresponding to three successive levels of HEALPix. Each tessera has a ray at its centre. The big bold tessera represents the solid angle $\Delta\theta_{\ell}$ of a single ray at level $\ell$. The four intermediate tesserae represent the solid angles of its child-rays at level $\ell+1$. And the sixteen smallest tesserae represent the solid angles of its grandchild-rays at level $\ell+2$. }
\label{fig: healpixtessel}
\end{figure}

The \texttt{NESTED} version of HEALPix allocates identifiers to rays according to the scheme illustrated in Fig. \ref{fig: healpixtessel}, which shows a small patch on the celestial sphere. Here the mother-ray of ray $m$ has ID
\begin{equation}
m'={\rm INT}\left(\frac{m}{4}\right)
\end{equation}
and the four child-rays of ray $m$ have IDs
\begin{equation}
m''=4m+n,\ {\rm with}\  n=0,1,2,3
\end{equation}
Here $n=0$ corresponds to the child-ray to the South, $n=1$ corresponds to the child-ray to the West, $n=2$ corresponds to the child-ray to the East and $n=3$ corresponds to the child-ray to the North. This numbering scheme makes it very simple to trace a ray back to the star at the origin, provided you know the ID of the ray \emph{and} its level.

\subsection{Ray rotation}
\label{rayrotation}

To avoid numerical artifacts due discretization in the directions of rays, we rotate the ensemble of HEALPix rays through three random angles (about the $z-$, $x'-$, and $z''-$axes), each time we build the ensemble. This process is necessary to prevent the formation of artificial corrugations in the ionization front, as explained in detail in Krumholz et al. (2007).


\subsection{Setting temperatures}
\label{givetemp}

At each evaluation point we estimate the value of $I(r_j)$ using Eqn. (\ref{sumint}). If the calculation returns $I(r_j)<I_{_{\rm MAX}}$, then the evaluation point lies in the interior of the {\HIIR} . Therefore, some of the particles that belong to its neighbour list should be ionized. We flag a particle as ionized, if it passes the following two tests: (a) its coordinates are inside the solid angle of the ray; and (b) its distance from the star is smaller than the distance of the evaluation point ($r_j$). If a particle is ionized, it is given a temperature $T=T_{\rm i}$. Otherwise it remains neutral with temperature $T=T_{\rm n}$. Fig.{\ref{fig: givet} illustrates the location of the neighbours of evaluation point $j$, and the ones which pass these tests.

\begin{figure}[h]
\centering
\includegraphics[width=0.4\textwidth]{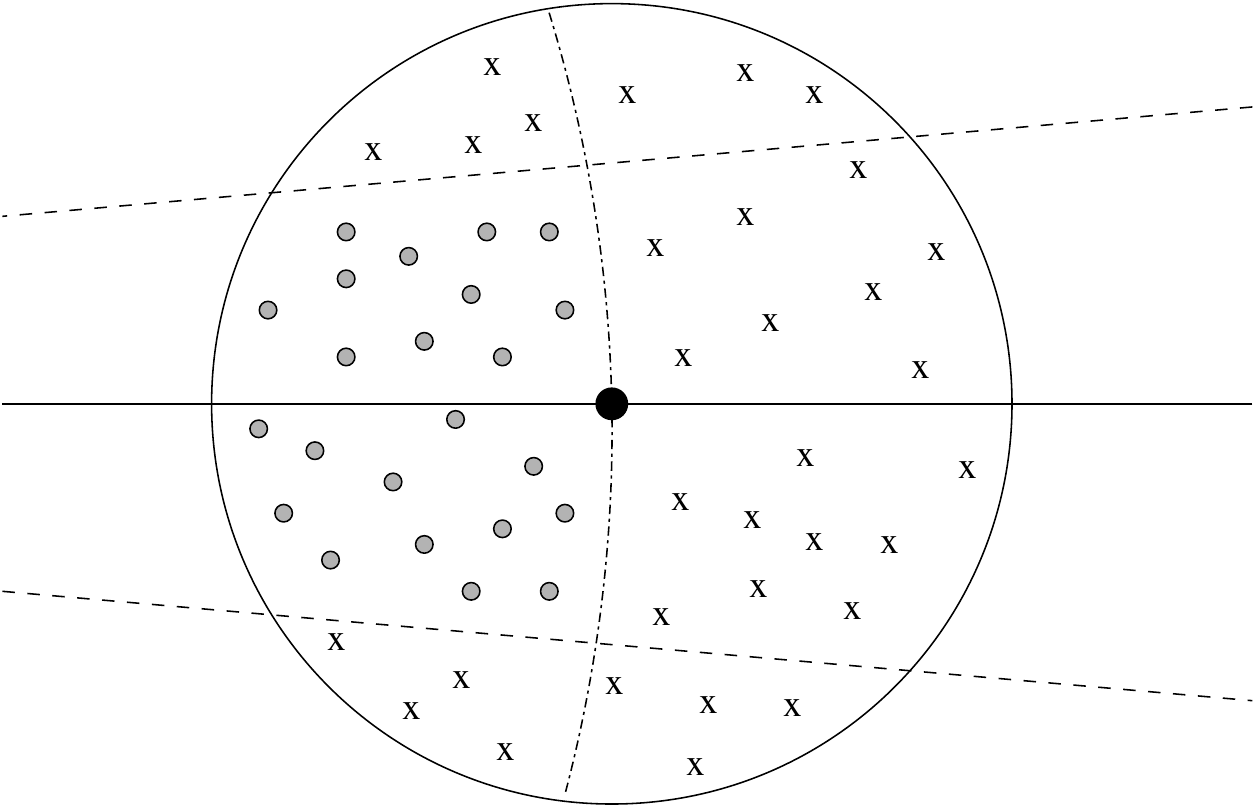}
\caption{ This figure shows a single ray (solid line) along with the boundaries of its solid angle (dash lines); the star lies on the left hand side and the ionization front on the right. The big black dot is the evaluation point $j$, and the circle is its smoothing region. The dot dashed line shows the position where the distance from the star is the same as that of the evaluation point. Small grey dots are particles which are flagged as being ionized, and the crosses are particles which are flagged as being neutral.}
\label{fig: givet}
\end{figure}


\subsection{Location of the ionization front}
\label{locationif}

We store in memory the value of the integral in Eqn. (\ref{sumint}) and the coordinates of the evaluation point at the end of a given ray segment. The distance $R_{_{\rm IF}}$ of the ionization front from the star along a given ray is given by the condition $I\left(R_{_{\rm IF}}\right)\,=\,I_{_{\rm MAX}}$. It is determined using a binary chop algorithm between the last two evaluation points $r_{j-1}$ and $r_j$ for which $I\left(r_{j-1}\right)<I_{_{\rm MAX}}<I\left(r_j\right)$.The binary chop algorithm is stopped once the adjustment to $r_j$ becomes smaller than $10^{-3}h_{j-1}$.

\begin{figure}[h]
\centering
\includegraphics[width=0.4\textwidth]{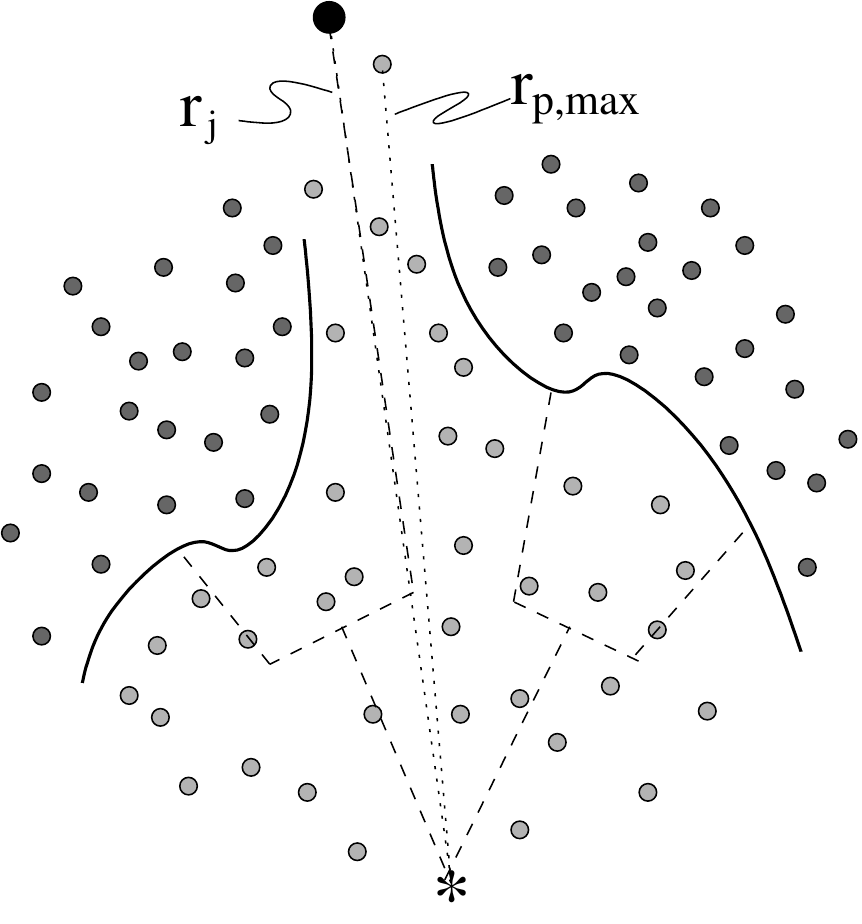}
\caption{ This figure shows how a ray penetrates the borders of the cloud. Once it finds no more neutral material to ionize along its line of sight, it is characterized as \emph{open}.}
\label{fig: openray}
\end{figure}

If the material of the cloud along the line of sight of a ray is sufficiently rarefied, $I\left(r\right)$ may never reach $I_{_{\rm MAX}}$ within the computational domain. We therefore estimate the maximum distance $r_{p,{\rm max}}$ of all gas particles and if $r_j>r_{p,{\rm max}}$ we stop calculating the sum.
Any ray that satisfies this condition is characterized as \emph{open} and the position of the ionization front is not determined along such rays (see Fig. \ref{fig: openray}).


\subsection{Temperature smoothing}
\label{tempsmoothing}

It is not feasible to resolve the ionization front in a standard SPH simulation of an evolving {\HIIR} (see Appendix \ref{ap.tempsmoothing}). Consequently we have to smooth the temperature gradient artificially across the ionization front. To demonstrate why this is necessary, we model a uniform density spherical cloud consisting of $10^6$ pre-settled SPH particles. The total mass of the cloud is $M\,=\,1000{\rm M_{\odot}}$, and the radius is $R\,=\,1{\rm pc}$. At the centre of the cloud there is a single star emitting $\dot{\cal{N}}_{_{\rm LyC}}\,=\,10^{49}$ ionizing photons per second. There are no gravitational forces. We simulate this system in two ways: in Case A, the temperature gradient is unsmoothed; and in Case B, the temperature gradient is smoothed.

In Case A, the temperatures of the SPH particles are set as described in \S\ref{givetemp}, and therefore there is an abrupt decrease in temperature across the ionization front. The simulation is terminated once the ionization front reaches the edge of the cloud. In Fig. \ref{fig.hnosm} the smoothing lengths of all SPH particles are plotted against their distance from the ionizing source at time $t\,=\,0.14\,{\rm Myr}$. Due to the abrupt temperature and pressure change across the ionization front, a gap is formed between the ionized and neutral regions. This gap prevents the ionization of new material from the neutral region during the expansion, and therefore the mass of the {\HIIR} does not increase (see Fig. \ref{fig.masscases}). This in turn alters the evolution of the radius of the {\HIIR}, as shown in Fig. \ref{fig.radiuscases}.

\begin{figure}
   \centering
   \includegraphics[width=0.4\textwidth]{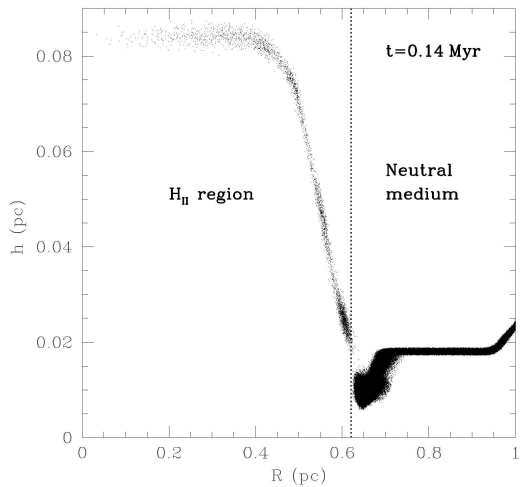}
      \caption{ The smoothing length of all ($10^6$) particles are plotted as a function of their distance from the star at $t\,=\,0.14\,{\rm Myr}$. Each point represents an individual SPH particle. The ionization front is located at $r_{_{\rm IF}}\,\sim\,0.63\,{\rm pc}$ away from the star. A gap between the ionized and neutral regimes is formed due to the discontinuous transition in temperature (Case A).
              }
         \label{fig.hnosm}
   \end{figure}

\begin{figure}
   \centering
   \includegraphics[width=0.4\textwidth]{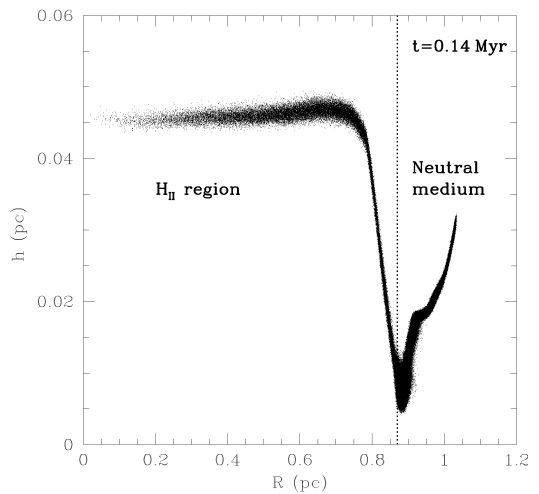}
      \caption{ The smoothing lengths of the particles are plotted as a function of their distance from the star, for the Case B. Despite that the snapshot is taken at the same time as in Fig. \ref{fig.hnosm} ($t\,=\,0.14\,{\rm Myr}$) the position of the ionization front is placed further, at $r_{_{\rm IF}}\,\sim\,0.87\,{\rm pc}$. The transition between the two regimes is continuous now.
              }
         \label{fig.hwism}
   \end{figure}

\begin{figure}
   \centering
   \includegraphics[width=0.4\textwidth]{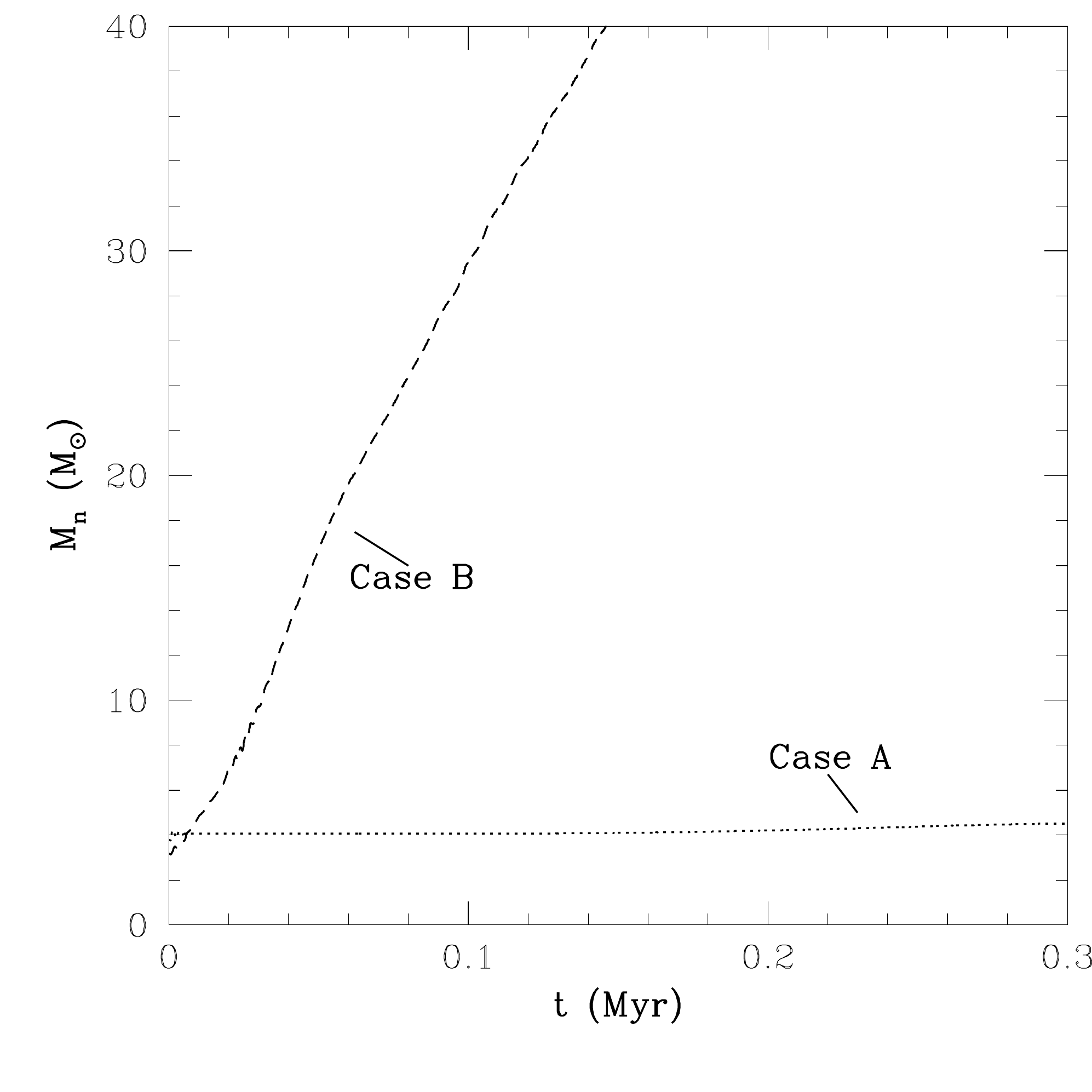}
      \caption{ Growth of mass of the {\HIIR} against time. In both cases we measure the mass of the gas that is completely ionized ($T_{\rm i}\,=\,10^4\,{\rm K}$). Clearly in Case B new mass is ionized while in Case A the mass is almost constant.
              }
         \label{fig.masscases}
   \end{figure}

\begin{figure}
   \centering
   \includegraphics[width=0.4\textwidth]{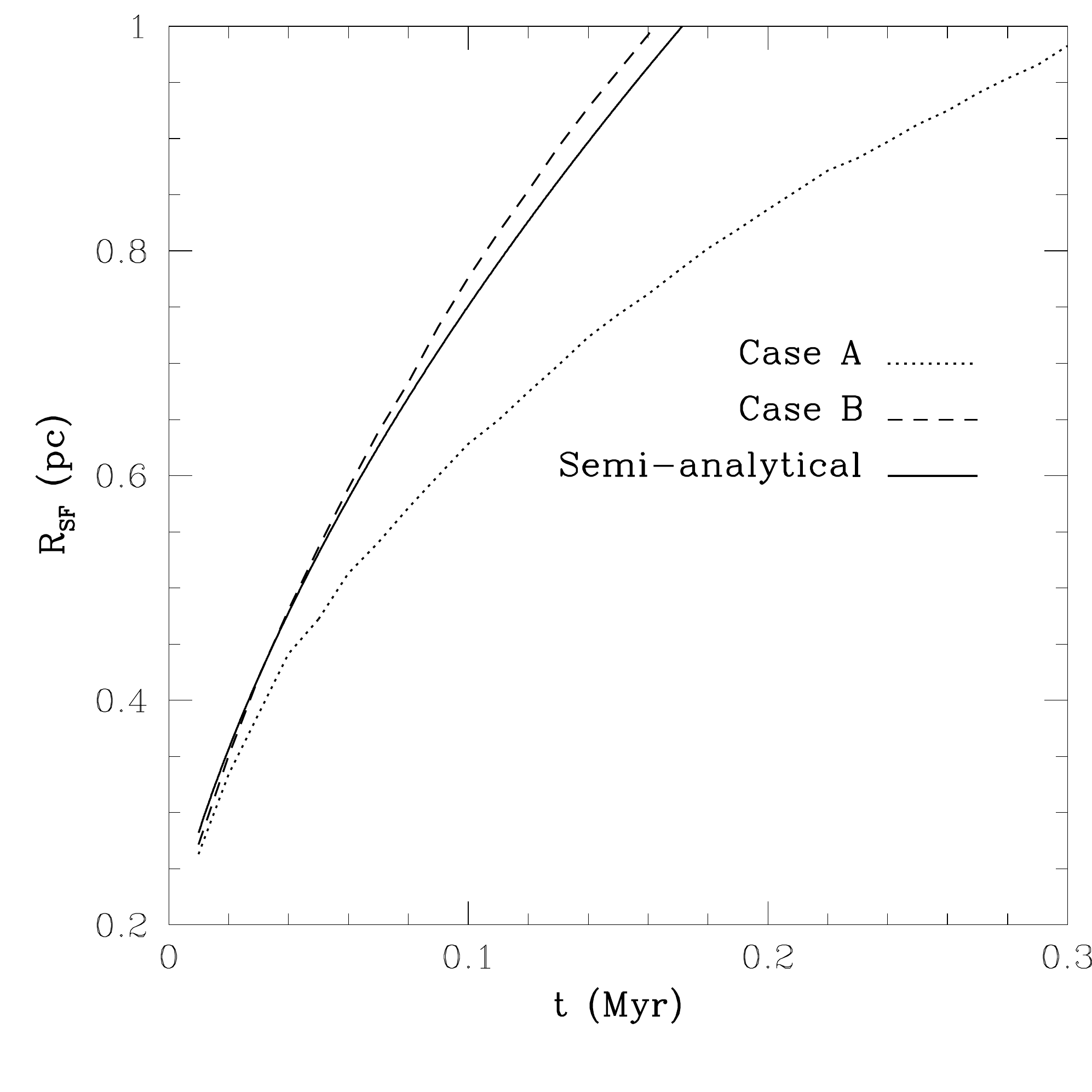}
      \caption{ The radius of the shell is plotted against time, for both cases A and B along with the semi-analytical solution (Eqn. \ref{semi_ode}). The evolution of the {\HIIR} in Case A is in substantial disagreement, while in Case B the {\HIIR} expands according to Eqn. (\ref{semi_ode}).
              }
         \label{fig.radiuscases}
   \end{figure}

In Case B, the temperature gradient across the ionization front is smoothed. Well inside the {\HIIR}, and well outside the ionization front, we set the temperatures of the SPH particles as described in \S\ref{givetemp}. However, in a layer of thickness $2h_{_{\rm IF}}$ across the ionization front we linearly interpolate between the temperatures in the ionized and neutral regimes, i.e.
\begin{eqnarray}\nonumber
T(r)&=&\frac{1}{2}\left(T_{\rm n}+T_{\rm i}\right)+\frac{r-r_{_{\rm IF}}}{2h_{_{\rm IF}}}\left(T_{\rm n}-T_{\rm i}\right),\;|r-r_{_{\rm IF}}|<h_{_{\rm IF}}.\\\label{eqn.tempsmooth}
\end{eqnarray}
Here $h_{_{\rm IF}}$ is the smoothing length of the local evaluation point on the ionization front, and $r_{_{\rm IF}}$ is its radius. In this case, the transition between the two temperature regimes is continuous (see Fig. \ref{fig.hwism}). Crucially, this allows new neutral material to be ionized (see Fig. \ref{fig.masscases}), and the expansion of the {\HIIR} now follows the analytical solution (Eqn. \ref{eq.spitzersolution}) rather closely.

Several other authors have encountered the same problem in the past. Kessel-Deynet \& Burkert (2000) smooth the ionization fraction over a distance of the order of one local smoothing length at the position where the ionization front is located. Krumholz et al. (2007) considers the same problem and constructs a region in which a similar procedure is followed.

\subsection{Computational cost}

The computational cost of locating the ionization front scales as ${\cal N_{\rm i}}\log\left({\cal N_{\rm i}}\right)$, where ${\cal N_{\rm i}}$ is the number of ionized SPH particles. This is demonstrated in Figure \ref{fig.cost}, which shows the CPU time, $t_{_{\rm iSUB}}$, taken by the ionization subroutine, as a function of ${\cal N_{\rm i}}$, for a sequence of simulations performed with different total numbers of SPH particles, ${\cal N}_{_{\rm TOT}}$; these are the same simulations as those used to test convergence in Section \ref{sec.hydro}.

\begin{figure}
   \centering
   \includegraphics[width=0.4\textwidth]{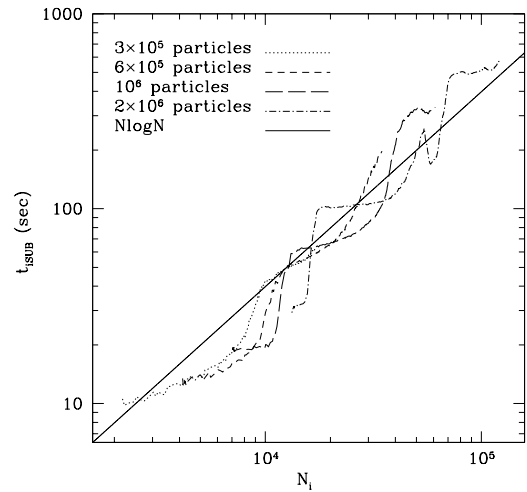}
      \caption{  This figure shows how the algorithm scales as the number of ionized particles increases. The abrupt changes in $t_{_{\rm iSUB}}$ occur when the level of refinement in HEALPix, $\ell$, increases by one. The discrepancies in $t_{_{\rm iSUB}}$ for specific ${\cal N_{\rm i}}$ are due to asymetrical branches of the families of rays in each simulation.}
         \label{fig.cost}
   \end{figure}

Abrupt increases in $t_{_{\rm iSUB}}$ occur when the level of refinement in HEALPix, $\ell$, increases by one, in response to the resolution criterion described in Section \ref{raysplitting}. Apart from this we see that the scaling fits ${\cal N_{\rm i}}\log\left({\cal N_{\rm i}}\right)$ very closely, as might be expected.

In this sequence of simulations, the overall cost of the ionization subroutine, over the entire duration of the simulation, scales as ${\cal N_{_{\rm SPH}}}\log\left({\cal N_{_{\rm SPH}}}\right)$, where ${\cal N}_{_{\rm SPH}}$ is the total number of SPH particles. 


%

\section{Applications}
\label{sec.applications}

In order to demonstrate the capabilities of the new algorithm, we present the results of four different applications. (i) In Section \ref{sec.hydro} we revisit the simulation described in Section \ref{tempsmoothing}, in which a spherically symmetric {\HIIR} expands inside a spherically symmetric, uniform-density, non--self-gravitating cloud. In Section \ref{tempsmoothing} this simulation was used to establish the need for temperature smoothing. Here we use it to explore the effect of changing the resolution by varying the total number of SPH particles. We show that with $\ga 300,000$ SPH particles, the simulations are well converged, both with one another, and with the analytic approximation derived by Spitzer (Eqn.$\,$\ref{eq.spitzersolution}). (ii) In Section \ref{sec:grav}, we repeat this simulation with a much more massive cloud, and include the effect of self-gravity. The evolution is now well described by the semi-analytic solution derived in Section \ref{semi-analytic} (see Eqn. \ref{semi_ode}). (iii) In Section \ref{sec.rocket}, we consider the same configuration as in Section \ref{sec.hydro}, except that the ionizing star is now initially located towards one edge of the cloud, and the self-gravity of the gas is neglected. As a consequence of the intrinsic asymmetry in the initial conditions, the {\HIIR} bursts out of the cloud on one side, and the remainder of the cloud on the other side is accelerated away by the rocket mechanism (Oort \& Spitzer 1955), and then ionized. (iv) In Section \ref{sec.rdi} we consider a much lower-mass, uniform-density, self-gravitating cloud which is overrun by an {\HIIR}. The ionization front propagating into the cloud is preceded by a shock front, which compresses the cloud,but does not trigger collapse.

In all simulations, the SPH particles are initially positioned randomly, and then settled to produce a uniform-density ``glass''; the temperature of the ionized gas is set to $T_{\rm i}=10^4\,{\rm K}$ (except in the boundary layer, where the temperature is smoothed to values between these two extremes, as described in Section \ref{tempsmoothing}); the recombination coefficient into excited states only is taken to be $\alpha_{_{\rm B}}=2.7\times 10^{-13}\,{\rm cm}^3\,{\rm s}^{-1}$; and the volume occupied by the ionizing star is neglected.

The parameters of all simulations are listed in Table \ref{table.param}. In this table $M$ is the mass of the cloud, $R$ is its radius, $T_{\rm n}$ is the temperature of the neutral medium, $X$ and $Y$ are the fractions of hydrogen and helium respectively, $\mu_{\rm n}$ and $\mu_{\rm i}$ are the mean molecular weights of the neutral medium and the ionized medium respectively, $\dot{\cal N}_{_{\rm LyC}}$ is the photon rate emission of the source, $D$ is the distance of the source from the centre of the cloud, $R_{_{\rm St}}$ is the initial Str{\o}mgren radius and ${\cal N}_{_{\rm SPH}}$ is the number of SPH particles used in each simulation.


\begin{table*}
\caption{This table is a summary of the parameters used in all applications.}             
\label{table.param}      
\centering          
\begin{tabular}{c c c c c c c c c c c c c }  
\hline\hline       

Application & $M\,({\rm M_{\odot}})$ & $R\,({\rm pc})$ & $T_{\rm n}\,({\rm K})$ & $X$ & $Y$ & $\mu_{\rm n}$ & $\mu_{\rm i}$ & ${\dot {\cal N}}_{_{\rm LyC}}\,({\rm s^{-1}})$ & $D\,({\rm pc})$ & $R_{_{\rm St}}\,({\rm pc})$ & Self-gravity & ${\cal N}_{_{\rm SPH}}$ \\
\hline                    
\S\ref{tempsmoothing} & 1000             & 1    & 10  &   0.7 & 0.3 & 2.35 & 0.678 &  $10^{49}$           &  0   &  0.189 & OFF & $10^6$\\
\S\ref{sec.hydro}     & 1000             & 1    & 10  &   0.7 & 0.3 & 2.35 & 0.678 &  $10^{49}$           &  0   &  0.189 & OFF & $3,6,10,20\times 10^5$\\  
\S\ref{sec:grav}      & $1.5\times 10^5$ & 14.6 & 10  &   0.7 & 0.3 & 2.35 & 0.678 &  $10^{49}$           &  0   &  1.43  & ON  & $10^7$\\
\S\ref{sec.rocket}    & 300              & 1    & 100 &   0.7 & 0.3 & 2.35 & 0.678 &  $10^{49}$           &  0.4 &  0.42  & OFF & $10^6$\\
\S\ref{sec.rdi}       & 20               & 0.5  & 100 &   1   & 0   & 1    & 0.5   &  $3.2\times 10^{48}$ &  3.5 &  --    & ON  & $3\times 10^5$\\
\hline                  
\end{tabular}
\end{table*}

\subsection{Spherically symmetric expansion of an {\HIIR} in a non--self-gravitating uniform-density cloud}
\label{sec.hydro}

In this application, we create a uniform-density spherical cloud having mass $M\,=\,1000\,{\rm M_{\odot}}$, initial radius $R\,=\,1\,{\rm pc}$, and hence initial density $\rho_{\rm n}=1.6\times 10^{-20}\,{\rm g}\,{\rm cm}^{-3}$. An ionizing source is placed at the centre of the cloud, and emits ionizing photons at a constant rate $\dot{\cal N}_{_{\rm LyC}}\,=\,10^{49}\,{\rm s}^{-1}$. Using Eqn. (\ref{eq.stradius}), the initial Str{\o}mgren radius is $R_{_{\rm St}}\,=\,0.189\,{\rm pc}$. The neutral gas is assumed to be at $T_{\rm n}=10\,{\rm K}$. There are no gravitational forces.

In order to establish that our code is converged, we evolve this configuration using different numbers of SPH particles: ${\cal N}_{_{\rm SPH}}=3\times10^5,\;6\times10^5,\;10^6,\;{\rm and}\;2\times10^6$. We terminate the simulations when the ionization front reaches the edge of the cloud. In Fig. \ref{fig.ifall} we plot the average radius of the ionization front, $R_{_{\rm IF}}$, against time for all four simulations. We see that the curves converge for ${\cal N}_{_{\rm SPH}}\ga 10^6$.

\begin{figure}
   \centering
   \includegraphics[width=0.4\textwidth]{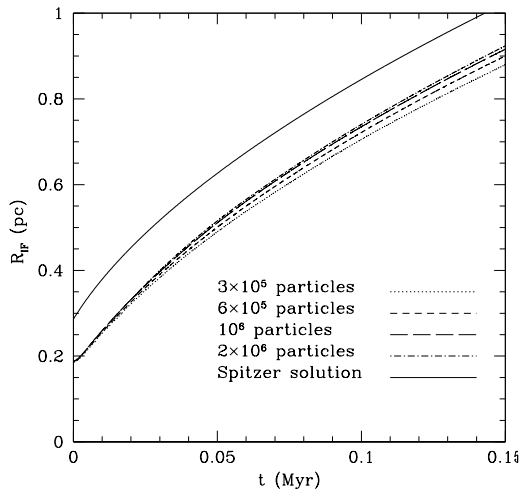}
      \caption{ The radius of the ionization front against time, for all four simulations in Section \ref{sec.hydro}. The simulations are terminated when the ionization front reaches the edge of the cloud. The solid line is the Spitzer solution, displaced upwards by $\Delta R_{_{\rm IF}}=0.1\,{\rm pc}$ to avoid confusion. Convergence is achieved for ${\cal N}_{_{\rm SPH}}\stackrel{>}{\sim}10^6$.}
         \label{fig.ifall}
   \end{figure}

In Fig. \ref{fig.ifsf} we plot the radius of the ionization front, $R_{_{\rm IF}}$, and the radius of the shock front, $R_{_{\rm SF}}$, against time, for the simulation performed with ${\cal N}_{_{\rm SPH}}=2\times10^6$ SPH particles. We determine the radius of the shock front by finding the most distant particle from the source which has radial outward velocity $v_{_{\rm r}}>0.1\,{\rm km}\,{\rm s^{-1}}$ and density $\rho>1.1\rho_{\rm n}$. The second condition is necessary because particles near the edge of the cloud move outwards due to the thermal pressure gradient there, long before they are overrun by the shock front. The factor $1.1$ is to accommodate numerical noise in the SPH estimate of the density of a particle.

In general, we find that the Spitzer solution (Eqn. \ref{eq.spitzersolution}) predicts the radius of the ionization front well, whereas the semi-analytic solution (Eqn. \ref{semi_ode}) predicts the position of the shock front. The advantage of the semi-analytic solution is that it can also be used to treat situations in which the self-gravity of the gas is important, as we show in the next application.

\begin{figure}
   \centering
   \includegraphics[width=0.4\textwidth]{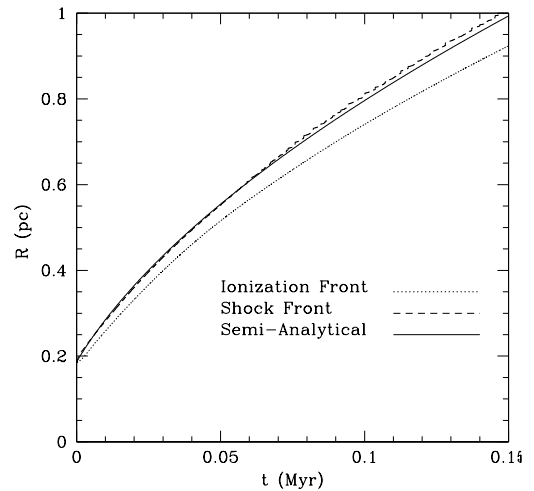}
      \caption{ The radii of the ionization front and the shock front against time, for the simulation of Section \ref{sec.hydro} performed with ${\cal N}_{_{\rm SPH}}=2\times10^6$ SPH particles. The solid line is the semi-analytic solution described in Sec.\ref{semi-analytic}.}
         \label{fig.ifsf}
   \end{figure}

\subsection{Spherically symmetric expansion of an {\HIIR} in a self-gravitating uniform-density cloud}
\label{sec:grav}

In this application, we simulate a much larger, uniform-density spherical cloud, having mass $M\,=\,1.5\times 10^5\,{\rm M_{\odot}}$, initial radius $R\,=\,14.6\,{\rm pc}$, and hence initial density $\rho_{\rm n}=7.63\times10^{-22}\,{\rm g}\,{\rm cm}^{-3}$. An ionizing source is placed at the centre of the cloud, and emits ionizing photons at a constant rate $\dot{\cal N}_{_{\rm LyC}}\,=\,10^{49}\,{\rm s}^{-1}$. Using Eqn. (\ref{eq.stradius}), the initial Str{\o}mgren radius is $R_{_{\rm St}}\,=\,1.43\,{\rm pc}$. The neutral gas is assumed to be at $T_{\rm n}=10\,{\rm K}$. We use $10^7$ SPH particles, and evolve the system for $4.5\,{\rm Myr}$.

In this simulation self-gravity is taken into account, but with the following two modifications. First, once the radius of the shock front has been determined (as described in \S\ref{sec.hydro}), we neglect the gravitational acceleration of all the SPH particles outside this radius. This is to prevent infall of the undisturbed neutral gas. Otherwise the outer parts of the cloud are already falling quite rapidly towards the centre by the time the shock front reaches them, and this complicates the interpretation of the dynamics. Secondly, we only take account of the radial component of the gravitational acceleration. We do this to suppress fragmentation of the shell, again in order to keep the dynamics of the shell as simple as possible (i.e. spherically symmetric).

In Fig. \ref{fig.semianl} we plot the position of the ionization front, the position of the shock front, and the semi-analytic approximation. We see that the position of the shock front agrees very well with the semi-analytic approximation.

\begin{figure}
   \centering
   \includegraphics[width=0.4\textwidth]{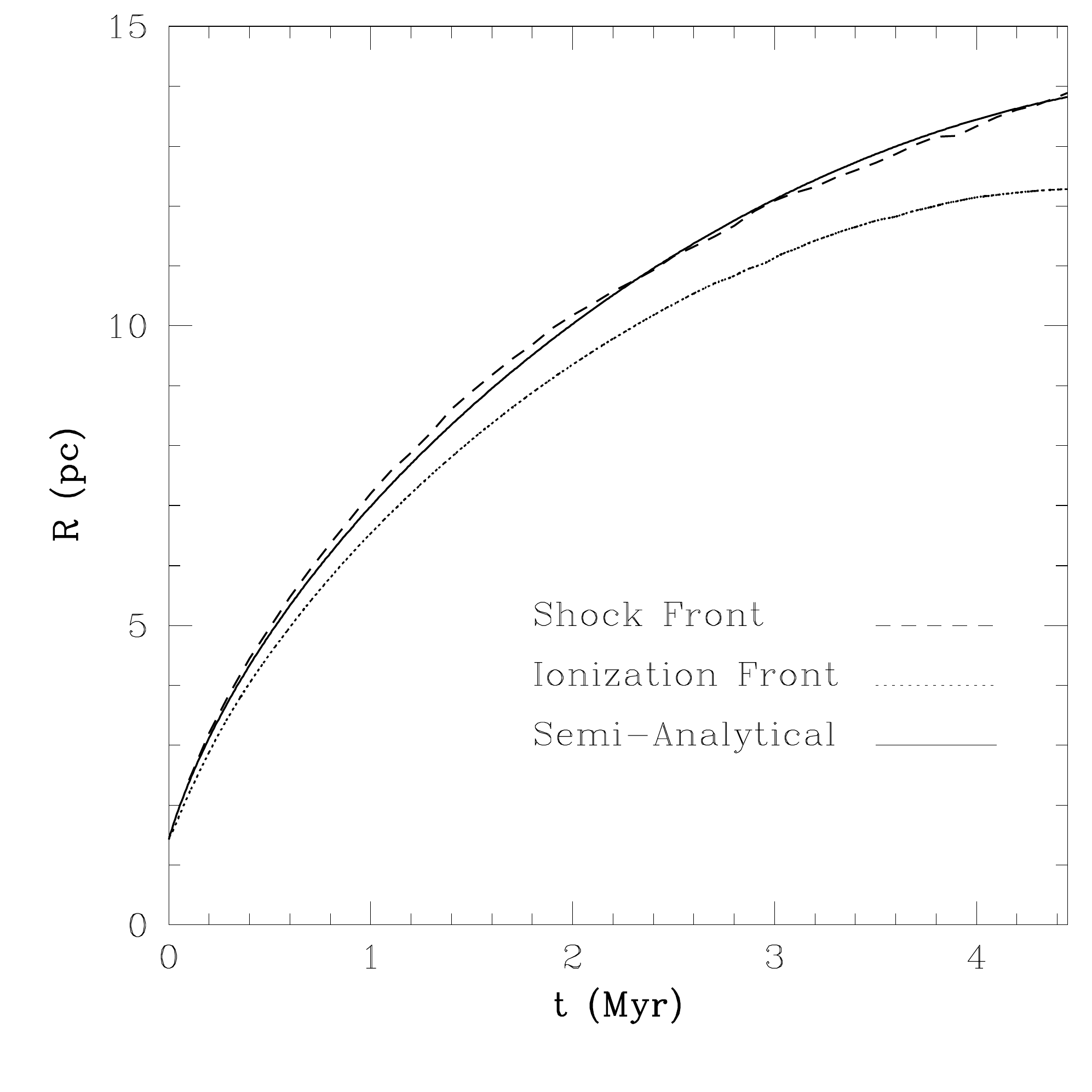}
      \caption{ This diagram shows the evolution of the shock front (dashed line) and the ionization front (dotted line) for the test described in \S\ref{sec:grav}. We plot Eqn. (\ref{semi_ode}) for comparison (solid line). The evolution of shock front agrees with the prediction of our semi-analytical approximation.
              }
         \label{fig.semianl}
   \end{figure}
\subsection{Off-centre expansion of an {\HIIR}}
\label{sec.rocket}

In this application, we consider a uniform-density spherical cloud, having mass $M=300\,{\rm M}_{\odot}$, initial radius $R=1\,{\rm pc}$, and hence initial density $\rho_{\rm n}\sim4.85\times10^{-21}\,{\rm g}\,{\rm cm}^{-3}$. An ionizing source is placed at distance $D\,=\,0.40\,{\rm pc}\,$ from the centre of the cloud, and emits ionizing photons at a constant rate $\dot{\cal N}_{_{\rm LyC}}\,=\,10^{49}\,{\rm s}^{-1}$; the initial Str{\o}mgren sphere therefore has radius $R_{\rm St}=0.42\,{\rm pc}$. The neutral gas is assumed to be at $T_{\rm n}=100\,{\rm K}$. The simulation uses ${\cal N}_{_{\rm SPH}}=3\times10^5$ particles. There are no gravitational forces. The simulation is terminated at $t=0.5\,{\rm Myr}$.

Fig. \ref{fig.rockseq} shows three stages in the evolution of the cloud. At $t\sim0.018\,{\rm Myr}$ (Fig. \ref{fig.rockseq}a), the ionization front breaks through the edge of the cloud on the left-hand (near) side. From this moment on, the ionized gas can stream away freely on this side. However, on the right-hand side the ionization front continues to drive an approximately parabolic shock front into the interior of the cloud. By $t\sim0.07\,{\rm Myr}$ (Fig. \ref{fig.rockseq}b), this shock has passed the centre of the cloud. At this stage the swept-up layer of neutral gas starts to break up into small ``cometary knots''. By $t\sim0.14\,{\rm Myr}$ (Fig. \ref{fig.rockseq}c), the shock front has opened up into a more hyperbolic shape, and has reached the right-hand edge of the cloud. The shocked neutral gas has spawned even more cometary knots. Fig. \ref{fig.rockdet} shows a close-up of a few of the knots at $t\sim 0.38\,{\rm Myr}$. At $t\sim 0.50\,{\rm Myr}$, about $6\,\%$ of the gas remains neutral. All this neutral gas is contained in hundreds of small cometary knots, which are steadily being ablated by the ionizing flux. 

\begin{figure*}
   \centering
   \includegraphics[width=0.48\textwidth,angle=-90]{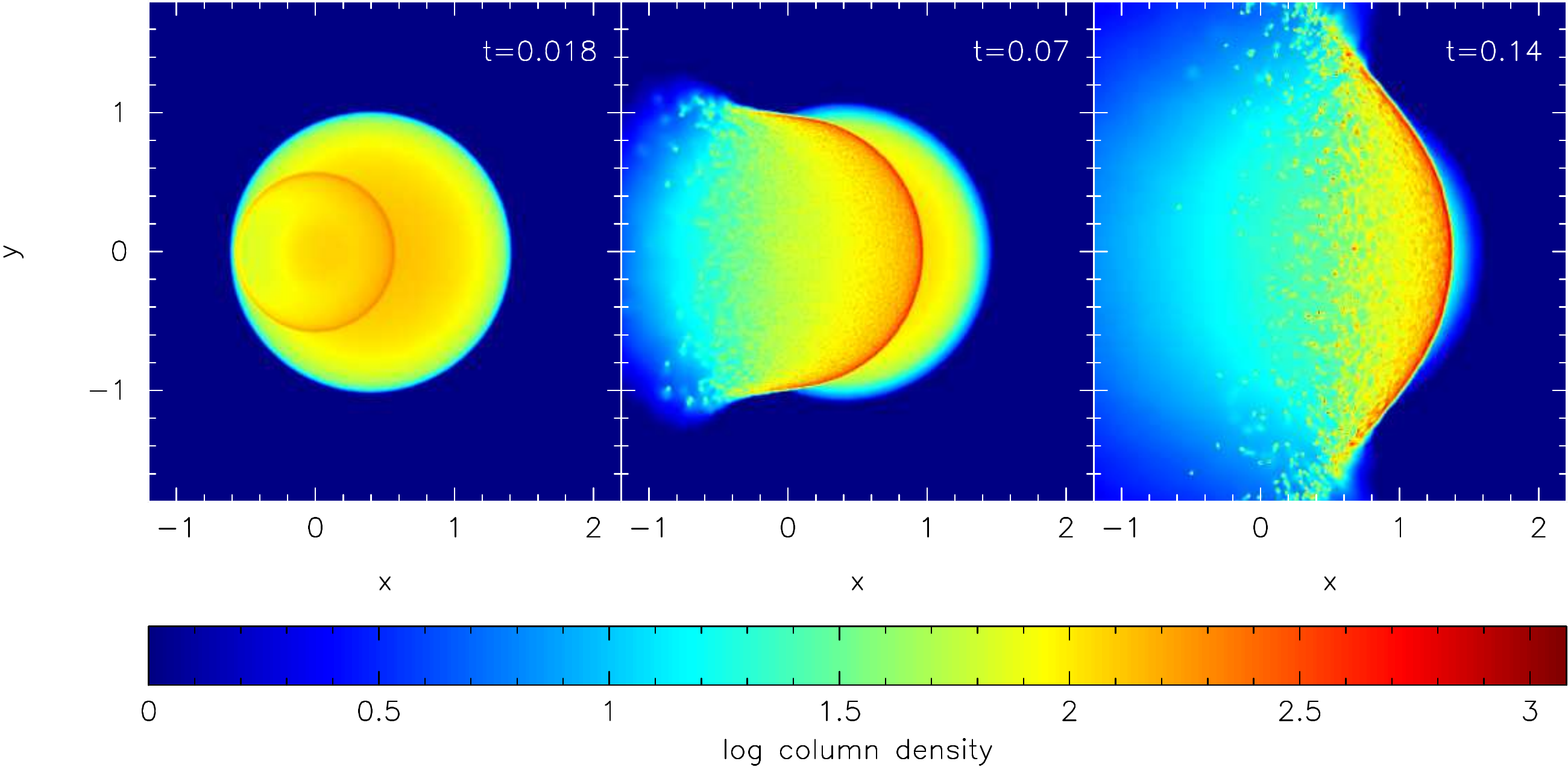}
      \caption{Column-density plots of the off-centre expansion of the {\HIIR}. Column-densities, $\Sigma$ are measured in ${\rm M}_\odot\,{\rm pc}^{-2}\;\left(\equiv 8.7\times 10^{19}\,{\rm H}_2{\rm cm}^{-2}\right)$. The spatial axes ($x$ and $y$) are labelled in parsecs, and times in Myr are given in the top right-hand corner of each frame.
              }
         \label{fig.rockseq}
   \end{figure*}

\begin{figure}
   \centering
   \includegraphics[width=0.5\textwidth,angle=-90]{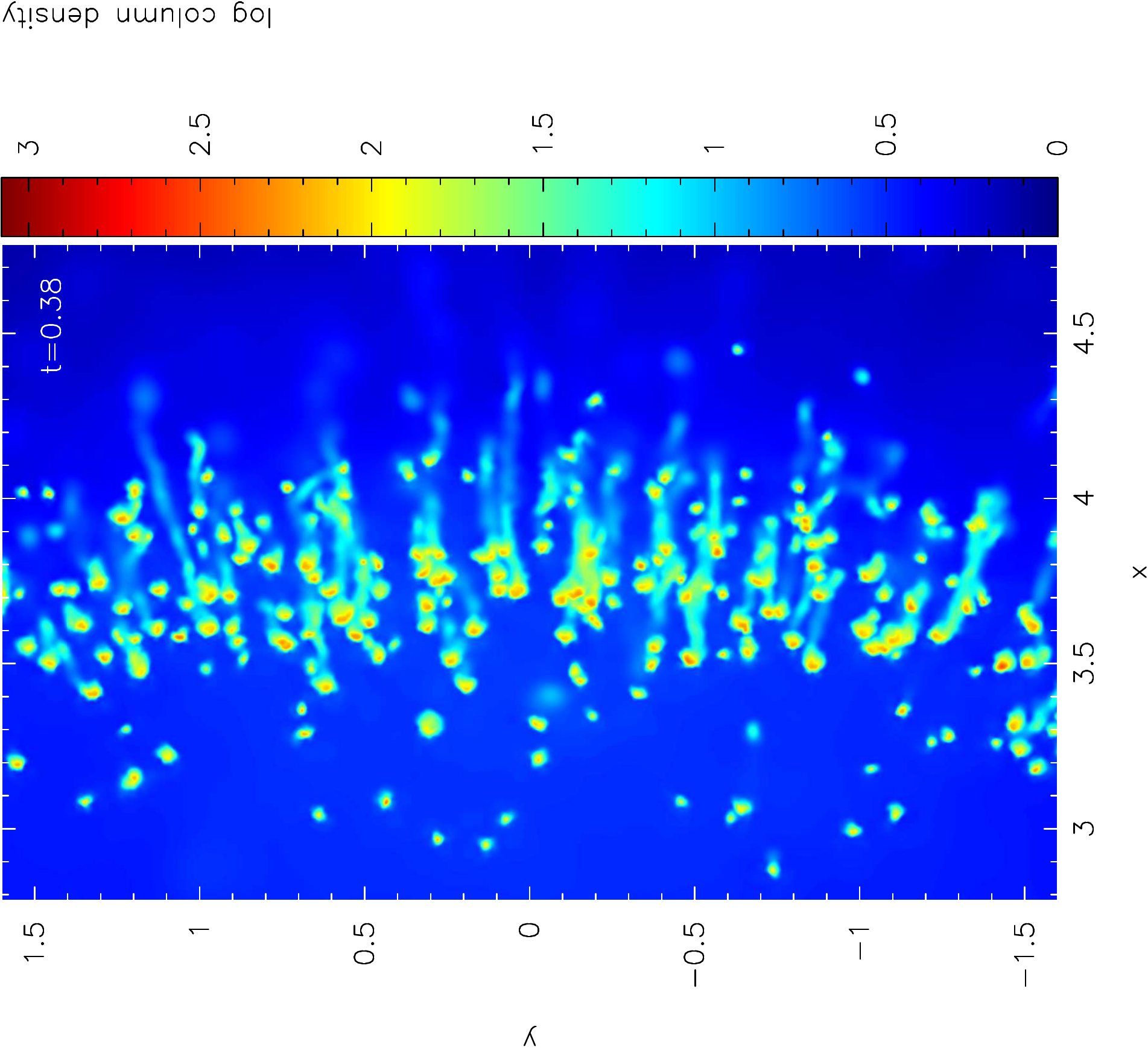}
      \caption{Detail of the off-centre expansion of the {\HIIR}. This column density plot shows a fraction of the hundred cometary knots formed at $t\sim 0.38\,{\rm Myr}$. Column density, $\Sigma$, is measured in ${\rm M}_\odot\,{\rm pc}^{-2}\;\left(\equiv 8.7\times 10^{19}\,{\rm H}_2{\rm cm}^{-2}\right)$. The spatial axes ($x$ and $y$) are labelled in parsecs.
              }
         \label{fig.rockdet}
   \end{figure}

We believe, but cannot unambiguously demonstrate, that the formation of these cometary knots is due to a combination of hydrodynamical instabilities. In the early stages, the shell should be prone to the Vishniac instability (Vishniac 1983); and later on, when all the gas has been swept up into the shell and is being accelerated, it should be prone to the Rayleigh-Taylor instability. Therefore we expect the shell to break up into small knots. However, the size of the knots in our simulation is just larger than the resolution limit of the code, i.e. individual knots contain $\sim 100\,$ SPH particles, and therefore have masses $\sim 0.1\,{\rm M}_\odot$ and diameters $\sim 0.03\,{\rm pc}$. It is not possible to establish the reality of the knots by performing convergence tests; if we increase the total number of SPH particles, ${\cal N}_{_{\rm TOT}}$, the masses of the knots decrease correspondingly. We note that these knots are very reminiscent of those seen in the Helix Nebula (O'Dell \& Handron 1996), and we will attempt to model this source in a future publication (cf. Capriotti 1973; O'Dell \& Burkert 1996; Capriotti \& Kendall 2006).

\subsection{Radiation Driven Compression}
\label{sec.rdi}

In this application, we consider a uniform-density spherical cloud, having mass $M=20\,{\rm M}_{\odot}$, initial radius $R=0.5\,{\rm pc}$, and hence initial density $\rho_{\rm n}\sim 2.6\times10^{-21}\,{\rm g}\,{\rm cm}^{-3}$. We place an ionizing source a distance $\,D=3.5\,{\rm pc}\,$ away from the centre of the cloud. The source emits ionizing photons at a constant rate $\dot{\cal N}_{_{\rm LyC}}\,=\,3.2\times 10^{48}\,{\rm s}^{-1}$. Hence the number-flux of ionizing photons incident on the near side of the cloud is $\Phi\,\sim\,2.97\times10^{9}\,{\rm cm}^{-2}\,{\rm s}^{-1}$. Since $\,D\gg R$, the rays are parallel and the flux is constant. In this one application we assume $X=1$ instead of $X=0.7$, i.e. pure hydrogen with $\mu_{\rm n}=1$ and $\mu_{\rm i}=0.5$. The neutral gas is assumed to be at $T_{\rm n}=100\,{\rm K}$. The sound speeds are therefore $c_{\rm n}=0.9\,{\rm km}\,{\rm s}^{-1}$ and $c_{\rm i}=12.8\,{\rm km}\,{\rm s}^{-1}$. The simulation uses ${\cal N}_{_{\rm SPH}}=3\times10^5$ particles, and self-gravity is taken into account. The simulation is terminated at $t=2.5\,{\rm Myr}$.

The above initial conditions are chosen to reproduce -- as closely as possible -- the initial conditions of the 2D model by Lefloch \& Lazareff (1994). We note that from the outset the cloud is over-pressured, and left to its own devices would simply disperse on a timescale of $\sim 1\,{\rm Myr}$. 

Fig. \ref{fig.rdiseq} shows the evolution of the cloud. The ionizing flux propagates upwards and rapidly boils off the outer layers on the near side of the cloud. At $t\sim 0.036\,{\rm Myr}$ (Fig. \ref{fig.rdiseq}a) a shock front starts to compress the remaining neutral gas. At the same time, the north hemisphere starts to expand due to the thermal pressure of the atomic hydrogen. At $t\sim 0.13\,{\rm Myr}$ (Fig. \ref{fig.rdiseq}b) a dense, prolate, approximately ellipsoidal core forms. By $t\sim 0.18\,{\rm Myr}$ (Fig. \ref{fig.rdiseq}c), the prolate core has semi-major axis $0.08\,{\rm pc}$ and semi-minor axis $0.02\,{\rm pc}$; its mass is $\sim 6\,{\rm M}_\odot$. It is therefore thermally sub-critical, and does not collapse to form a star. Instead it is steadily ablated by ionization. By $t\sim 0.21\,{\rm Myr}$ (Fig. \ref{fig.rdiseq}d), the remnants of the cloud start to develop a cometary tail. By $t\sim 2.5\,{\rm Myr}$ (not shown on Fig. \ref{fig.rdiseq}), the last vestiges of the cloud are ionized, and they are $\sim 28\,{\rm pc}$ from the ionizing star.

Fig. \ref{fig.rdimass} plots the total mass of neutral gas against time. The undulations seen at $t\sim 0.25\,{\rm Myr}$, and at $t\sim 0.5\,{\rm Myr}$ are acoustic oscillations, excited as the cloud responds to the increase in external pressure.

Gritschneder et al. (2009; hereafter G09) have also simulated radiatively driven compression. However, the cloud that G09 treat (with mass $96\,{\rm M}_\odot$, radius $1.6\,{\rm pc}$, and temperature $10\,{\rm K}$) is much more massive and much colder than the one we treat here ($20\,{\rm M}_\odot$, $0.5\,{\rm pc}$, $100\,{\rm K}$), and therefore it is much less resistant to compression and more prone to triggered gravitational collapse. Furthermore, in their first two simulations, G09 use a larger ionizing flux than we do. As a consequence, the clouds in their simulations are more strongly compressed than ours, particularly at the leading edge (i.e. the edge exposed directly to the ionizing flux), and are triggered into gravitational collapse. In contrast, our cloud is more mildly compressed, and evolves towards a centrally condensed configuration but never becomes gravitationally unstable.

\begin{figure*}
   \centering
   \includegraphics[width=0.45\textwidth,angle=-90]{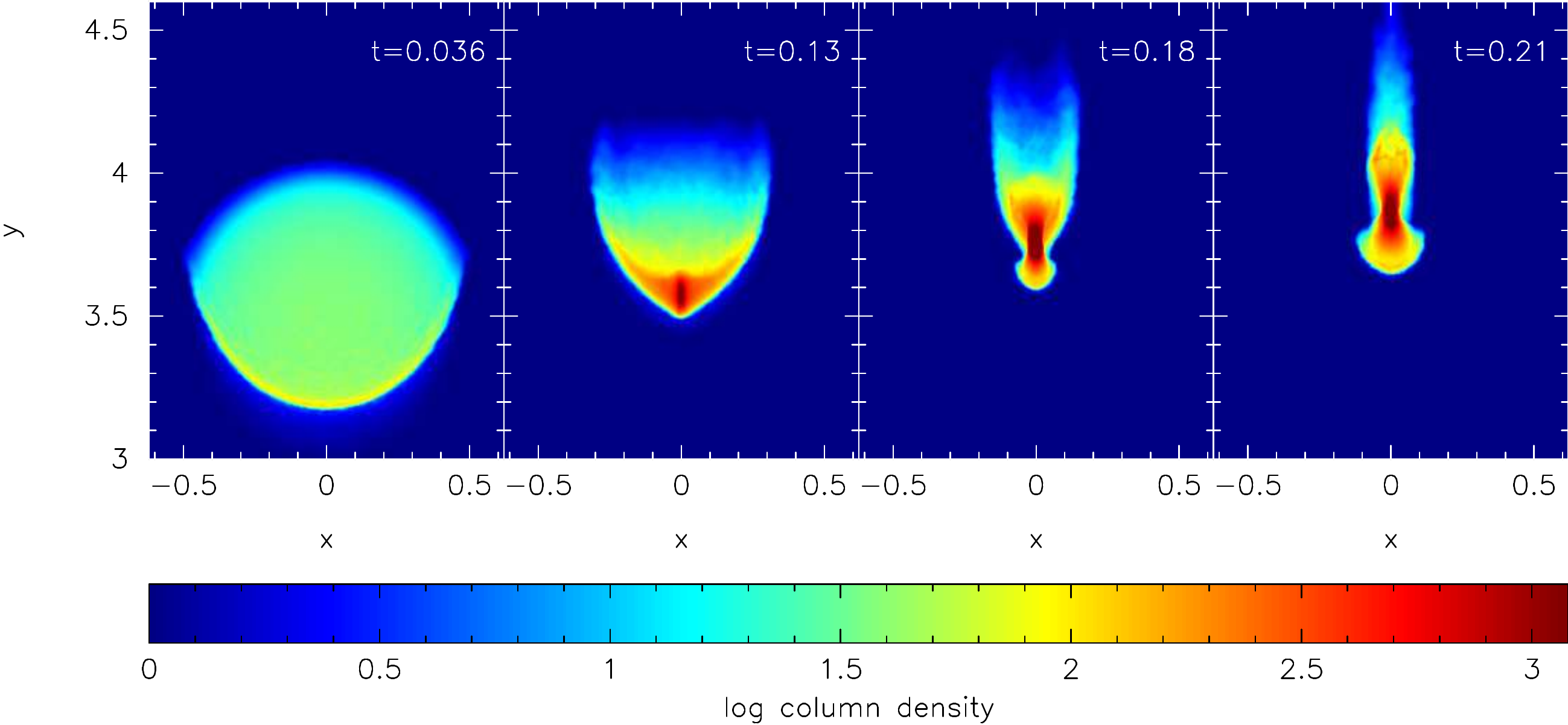}
      \caption{Column-density plots of radiation driven compression. Column-densities, $\Sigma$ are measured in ${\rm M}_\odot\,{\rm pc}^{-2}\;\left(\equiv 1.3\times 10^{20}\,{\rm cm}^{-2}\right)$. The spatial axes ($x$ and $y$) are labelled in parsecs, and times in Myr are given in the top right-hand corner of each frame.
              }
         \label{fig.rdiseq}
   \end{figure*}

\begin{figure}
   \centering
   \includegraphics[width=0.4\textwidth]{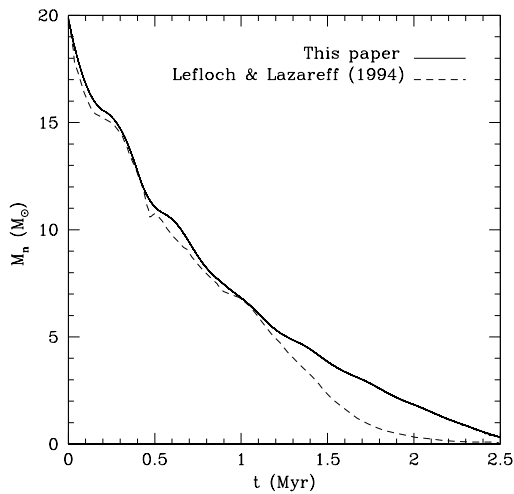}
      \caption{This diagram shows the evaporation in time of the neutral gas. The solid line represents the simulation discussed in \S\ref{sec.rdi}, whereas the dash line the simulation by Lefloch \& Lazareff (1994). The discrepancy between our simulation and the simulation by Lefloch \& Lazareff (1994) at time $t>1.1\,{\rm Myr}$ is due to the movement of the cloud further from the star, where the radiation flux drops in our model but stays constant in Lefloch \& Lazareff (1994) plane parallel model.
              }
         \label{fig.rdimass}
   \end{figure}

%

\section{Discussion \& Conclusions}
\label{sec.discussion}

We have introduced a new technique for treating the propagation of ionizing radiation in SPH simulations of self-gravitating gas dynamics. The method uses the HEALPix algorithm to tessellate the celestial sphere, and  solves the equation of ionization equilibrium along the rays associated with each tessera; rays are split hierarchically to produce greater resolution wherever it is required, i.e. to ensure that the resolution of the radiation transfer matches the resolution of the hydrodynamics, locally. This makes the algorithm very computationally efficient.

The algorithm has been incorporated into the new Smoothed Particle Hydrodynamics code SEREN (Hubber et al., in preparation), and tested against a number of known analytic and semi-analytic problems. These tests are presented here. The code follows the expansion of a spherically symmetric {\HIIR} in a non--self-gravitating gas (Spitzer solution); the expansion of a spherically symmetric {\HIIR} in a self-gravitating gas (new semi-analytic solution described in Sec.\ref{semi-analytic}); the rocket acceleration and subsequent ablation of a massive cloud irradiated by an ionizing star on one side (Oort \& Spitzer 1955); and the radiatively driven compression of a pre-existing dense core engulfed by an expanding {\HIIR}. The code will be used in future to explore the role of {\HIIR}s in triggering and regulating star formation, injecting turbulent energy into the interstellar medium, and eroding molecular clouds.

\begin{acknowledgements}

R.W. acknowledges support by the Human Resources and Mobility Programme of the European Community under contract MEIF-CT-2006-039802. The computations in this work were carried out on Merlin ARCCA SRIF-3 Cluster. The column density plots presented in this paper were prepared using SPLASH (Price 2007). 

\end{acknowledgements}
%

\appendix
\section{Resolution requirements to resolve the ionization front}
\label{ap.tempsmoothing}

Consider an {\HIIR} having uniform density $\rho_{\rm i}$. The SPH particles have a universal mass $m_{_{\rm SPH}}$. If each SPH particle has ${\cal N}_{_{\rm NEIB}}$ neighbours, then the diameter, $d$, of an SPH particle (i.e. the diameter of its smoothing kernel) is given by
\begin{eqnarray}\nonumber
\frac{\pi\,d^3\,\rho_{\rm i}}{6}&=&{\cal N}_{_{\rm NEIB}}\,m_{_{\rm SPH}}\,,\\
d&=&\left(\frac{6\,{\cal N}_{_{\rm NEIB}}\,m_{_{\rm SPH}}}{\pi\,\rho_{\rm i}}\right)^{1/3}\,.
\end{eqnarray}
Since $d$ is in effect the resolution of the simulation, the ionization front can only be resolved if $d\la \Delta R_{_{\rm IF}}$. Using Eqn. (\ref{eqn.ifthick}), this requirement reduces to
\begin{eqnarray}
m_{_{\rm SPH}}&\la&\frac{\pi\,(20)^3\,m^3}{6\,\dot{\cal N}_{_{\rm LyC}}\,\rho_{\rm i}^2\,\bar{\sigma}^3}\,.
\end{eqnarray}
Combining Eqns. (\ref{eq.solutionhiimass}) and (\ref{eq.solutionhiidensity}), we see that the mass of the {\HIIR} is given by
\begin{eqnarray}
M_{\rm i}&=&\frac{m^2\,\dot{\cal N}_{_{\rm LyC}}}{\alpha_{_{\rm B}}\,\rho_{\rm i}}\,.
\end{eqnarray}
The number of SPH particles required to model the {\HIIR} is therefore
\begin{eqnarray}
{\cal N}_{\rm i}&=&\frac{M_{\rm i}}{m_{_{\rm SPH}}}\\
&\ga&\frac{6\,\bar{\sigma}^{\,3}\,{\cal N}_{_{\rm NEIB}}\,\dot{\cal N}_{_{\rm LyC}}\,\rho_{\rm i}}{(20)^3\,\pi\,\alpha_{_{\rm B}}\,m}\;\\
&\ga&5\times10^{11}\,\left(\frac{{\cal N}_{_{\rm NEIB}}}{50}\right)\,\left(\frac{\dot{\cal N}_{_{\rm LyC}}}{10^{49}\,{\rm s}^{-1}}\right)\,\left(\frac{\rho_{\rm i}}{10^{-20}\,{\rm g}\,{\rm cm}^{-3}}\right)\,.
\end{eqnarray}
Evidently this is a prohibitive requirement on ${\cal N}_{\rm i}$, all the more so when one allows that the mass of neutral gas is likely to be even greater than the mass of ionized gas, and so this will require additional SPH particles.

\end{document}